\documentstyle[12pt,epsf]{article}
\begin{document}
\textheight 7.6in
\begin{flushright}
\mbox{BA-99-05}\\
\mbox{FERMILAB-Pub-99/007-T}\\
\mbox{January 1999} \\[0.5in]
\end{flushright}
\begin{center}
	{\large \bf Predicting quark and lepton masses and mixings\\[0.5in]}
 Carl H. Albright$^1$ \\ Department of Physics \\ Northern
Illinois University, DeKalb, IL 60115 \\
and Fermi National Accelerator Laboratory \\
P.O. Box 500, Batavia, IL 60510 \\[0.2in]
S.M. Barr$^2$ \\
Bartol Research Institute \\ University of Delaware \\
Newark, DE 19716\\[0.4in]
\end{center}


\begin{abstract}

A model is presented that fits the quark and lepton masses and mixings
wherein five dimensionless parameters and a phase account for fifteen 
dimensionless observables.  Among these are 
the Wolfenstein parameters $\rho$ and $\eta$, the $\nu_e-\nu_{\mu}$ 
and $\nu_e-\nu_{\tau}$ mixing angles which are predicted to be small and 
comparable while the $\nu_{\mu}-\nu_{\tau}$ mixing angle is predicted to be 
large. The model is based on supersymmetric $SO(10)$ with the form of the 
mass matrices motivated by simplicity at the level of grand
unification.\\[0.3in]
\end{abstract}
$\line(75,0){75}$\\
\noindent
$^1$E-mail: albright@fnal.gov; $^2$E-mail: smbarr@bartol.udel.edu\\
\thispagestyle{empty}
\newpage

In an earlier paper with Babu \cite{abb2}, the authors have proposed a model 
based on supersymmetric $SO(10)$ for the masses and mixings of the heavier 
two families of quarks and leptons.  One of the guiding principles in 
constructing the model was that the mass terms should be of a kind that 
can arise simply in $SO(10)$ with small representations of Higgs fields
\cite{ab1}.  Two critical points emerged 
from that work: namely, that it is quite natural in unified theories
for there to be flavor-asymmetric contributions to the fermion 
mass matrices, and that these can explain some of the puzzling 
features of the spectrum.  In particular, it was shown that a 
flavor-asymmetric term can account in a simple way for the apparent 
fact that the $\nu_\mu-\nu_\tau$ mixing is very large compared to $V_{cb}$.

The present work is an extension of that earlier model to include
the first family.  We find that it is only necessary to
introduce two new parameters, one of which has a complex phase,
in order to account for seven quantities that involve the first
family. In doing so, several interesting predictions result, 
including the values of the $\nu_e-\nu_{\mu}$ and $\nu_e- \nu_{\tau}$
mixing angles, and a relationship between the real and imaginary parts 
of $V_{ub}$, {\it i.e.}, between the Wolfenstein parameters $\rho$
and $\eta$. Altogether, as will be seen, the model leads to  nine
predictions.

\vspace{0.5cm}

\noindent
{\bf The mass matrices.}

\vspace{0.2cm}

The mass matrices proposed have the following forms:

\begin{equation}
\begin{array}{ll} 
U^0 = \left( \begin{array}{ccc}
0 & 0 & 0 \\
0 & 0 & \epsilon/3 \\
0 & -\epsilon/3 & 1 
\end{array} \right) M_U, &

D^0 = \left( \begin{array}{ccc}
0 & \delta & \delta' \\
\delta & 0 & \sigma + \epsilon/3 \\
\delta' & - \epsilon/3 & 1 
\end{array} \right) M_D \\

& \\

N^0 = \left( \begin{array}{ccc}
0 & 0 & 0 \\
0 & 0 & - \epsilon \\
0 & \epsilon & 1
\end{array} \right) M_U, &

L^0 = \left( \begin{array}{ccc}
0 & \delta & \delta' \\
\delta & 0 & -\epsilon \\
\delta' & \sigma + \epsilon & 1 
\end{array} \right) M_D
\end{array}
\end{equation}

\noindent
where $U$, $D$, $L$, and $N$ stand, respectively, for the
up-quark, down-quark, charged-lepton, and
the neutrino Dirac mass matrices. The zero superscripts stand 
here and throughout for quantities defined at the grand unification
(GUT) scale. The convention we use is that the left-handed fermions
multiply these matrices from the right, while the 
right-handed fermions multiply them from the left.
(Thus, it is the entries below the main diagonal that control the
quark and lepton mixings that are observed.)

It is immediately apparent that the matrix $U^0$ is rank-2,
corresponding to a massless $u$ quark. We imagine that some
small higher-order effects give $u$ a mass of several MeV. The fact that
$m_u$ must arise at higher order than $m_d$ and $m_e$ is not a flaw but
a qualitative
success of the model. The point is that a several-MeV $u$-quark mass
corresponds to $m_u^0/m_t^0 \approx 10^{-5}$. This is very much smaller 
than the corresponding ratios $m_d^0/m_b^0 \approx 10^{-3}$ and
$m_e^0/m_{\tau}^0 \approx 0.3 \times 10^{-3}$, and so it is reasonable
to suppose that it is a higher-order effect.

By field redefinitions one can make all the elements of the matrices
in Eq. (1) real except for $\delta'$. (There can also be a relative
phase {\it initially} between $\sigma$ and $\epsilon$, which was called
$\alpha$ in \cite{abb2}. After phase redefinitions, however, this parameter 
survives only in the entries $D_{23}^0$ and $L_{32}^0$, which can
be approximated by $\sigma + \frac{\epsilon}{3} \cos \alpha$ and $\sigma
+ \epsilon \cos \alpha$, respectively, since as will be seen later 
$\epsilon/\sigma$ is a small parameter. As $\cos \alpha$ only appears
in terms subleading in $\epsilon/\sigma$, it has only a few-percent effect on 
predictions. 
In \cite{abb2} it was found that the fits prefer $\alpha = 0$, so we will 
assume that value henceforth in this paper.) Effectively, then, 
the mass matrices in Eq. (1) 
contain five real dimensionless parameters and one phase: 
$(M_U/M_D)$, $\sigma$, $\epsilon$, $\delta$, $\left| \delta' \right|$,
and $\arg \delta'$.

We will first describe the kinds of $SO(10)$ operators and diagrams
that give rise to the entries shown in (1), and explain the
rationale for the those entries. Then 
we shall proceed to discuss the fit to the data and the
predictions of the model. 

\vspace{0.5cm}

\noindent
{\bf The origin and rationale of the matrices.}

\vspace{0.2cm}

The (33) entries of the mass matrices contribute to the masses
of the heaviest family of quarks and leptons. As is
true of most grand unified models of fermion masses, these
entries can come from the minimal kind of Yukawa term.
Since ours is an $SO(10)$ model, these entries
arise simply from the operator $({\bf 16}_3 {\bf 16}_3) {\bf 10}_H$.
Here and throughout, the subscript `$H$' stands for a Higgs
multiplet, and a numerical subscript stands for the family
number of a quark/lepton multiplet.  The parameters $M_U$ and $M_D$ 
in Eq. (1) come from the $SU(2)_L \times U(1)_Y$-breaking vacuum 
expectation values (VEVs) in ${\bf 5}({\bf 10}_H)$ and 
$\overline{{\bf 5}}({\bf 10}_H)$, respectively. 
We use (${\bf p}({\bf q})$ to represent a ${\bf p}$ multiplet of $SU(5)$
contained in a ${\bf q}$ multiplet of $SO(10)$.

While the masses of the third family satisfy the minimal $SU(5)$
(and $SO(10)$) relation $m_b^0 \cong m_{\tau}^0$,
the middle family does not satisfy the corresponding
$m_s^0 \cong m_{\mu}^0$ relation. As is well known, however, a good fit
is given by the Georgi-Jarlskog relation \cite{g-j}, $m_s^0 \cong m_{\mu}^0/3$.
Perhaps the simplest way to get a factor of 1/3 into the
elements of the mass matrices is through an adjoint Higgs multiplet,
${\bf 45}_H$. Such a Higgs multiplet is needed in any case for the breaking
of $SO(10)$ down to the standard model group. In the simplest
breaking scheme that naturally solves the ``doublet-triplet
splitting problem", the so-called Dimopoulos-Wilczek or
``missing VEV" mechanism \cite{d-w}, there is an adjoint Higgs multiplet whose
VEV is proportional to the $SO(10)$ generator $B-L$. This is in
fact just what is needed to get a relative factor of $1/3$
between entries in the quark and lepton mass matrices.

The lowest-dimension Yukawa operators that contains the adjoint
Higgs multiplet are products of the form 
${\bf 16}_i {\bf 16}_j {\bf 10}_H {\bf 45}_H$. There are
two ways to contract the $SO(10)$ indices, but it is easy to
show that if $\langle {\bf 45}_H \rangle \propto B-L$, then
only the {\it flavor-antisymmetric} piece contributes to the
mass matrices. This accounts for the form of the ``$\epsilon$" entries 
in Eq. (1). Note that these appear with a factor of $1/3$ in the
quark matrices and $1$ in the lepton matrices, reflecting the generator
$B-L$. One also sees that because the operator being discussed
contains the
same ${\bf 10}_H$ that appears in the Yukawa operator for the
(33) entries, the $\epsilon$ entries of $U^0$ and $N^0$
are $M_U/M_D$ times those of $D^0$ and $L^0$.

The operator ${\bf 16}_2 {\bf 16}_3 {\bf 10}_H {\bf 45}_H$
needed for the $\epsilon$ entries arises in a simple fashion by
integrating out a superheavy ({\it i.e.} GUT-scale)
family plus conjugate-family pair: ${\bf 16} + \overline{{\bf 16}}$ 
and is naturally of order $1/M_G$ as required.  This
is shown in Fig. 1. Such an operator could also arise from Planck-scale
physics, but it would then be too small. 

The next pieces of the mass matrices to be discussed are the ``$\sigma$" 
entries.  These have several rationales, and are in fact the key to the
whole model. We shall first explain how such entries arise, and then
why they are important for getting a realistic model.

The feature of the $\sigma$ entries that distinguishes them from
what is found in most published models is that they are neither
flavor-symmetric nor flavor-antisymmetric, but rather ``lopsided":
$\sigma$ appears in $D^0_{23}$ but not in $D^0_{32}$. 
Though it is not common to consider such lopsided
forms, there is really no reason from the point of view of
grand unification not to do so. In fact, it may well be the neglect of such
possibilities that has handicapped the search for realistic
quark and lepton mass matrices.

The $\sigma$ entries arise very simply from integrating out 
an $SO(10)$ vector ${\bf 10}$ of quarks and leptons
as shown in Fig. 2. The resulting operator is of the form
$[{\bf 16}_2 {\bf 16}_H]_{10} [{\bf 16}_3 {\bf 16}'_H]_{10}$.
The notation $[ ... ]_{10}$ indicates that what is contained
in the brackets is contracted into a ${\bf 10}$ of $SO(10)$
as a result of the virtual ${\bf 10}$ fermions appearing in Fig. 2.
The Higgs multiplet denoted ${\bf 16}_H$ is assumed to develop
a GUT-scale VEV in the ${\bf 1}({\bf 16})$ direction, while the other 
spinor Higgs, denoted ${\bf 16}'_H$,  gets a 
weak-scale $SU(2)_L$-doublet VEV in the
$\overline{{\bf 5}}({\bf 16})$ direction. Thus, in $SU(5)$
notation, the operator from Fig. 2 gives 
$(\overline{{\bf 5}}_2 {\bf 10}_3) \langle \overline{{\bf 5}}'_H
\rangle \langle {\bf 1_H} \rangle$.

One observes that since the left-handed down-quarks are in ${\bf 10}$'s
of $SU(5)$, not $\overline{{\bf 5}}$'s, this operator contributes
to $D_{23}$ and not $D_{32}$. On the other hand, since the left-handed
leptons are in $\overline{{\bf 5}}$'s, not ${\bf 10}$'s, the
operator contributes to $L_{32}$ and not $L_{23}$. 
As we shall see, this fact that the $\sigma$ entry
is transposed in $L$ relative to that in $D$ is very important
for understanding the large $\nu_{\mu} - \nu_{\tau}$ mixing
that may have been seen in the atmospheric neutrino data. This
transposition is a consequence of the group theory of $SU(5)$ which 
also implies that $D = L^T$ in the minimal $SU(5)$ model.

The upshot is that integrating out ${\bf 10}$'s of fermions
in $SO(10)$ can quite typically give just such lopsided entries as the 
$\sigma$ entries in Eq. (1). Note that the $\sigma$ entries only appear 
in $D$ and $L$, and not in $U$ and $N$, following from the
fact that ${\bf 16}'_H$ contains only $\overline{{\bf 5}}$ and not
${\bf 5}$ of $SU(5)$. 

Now, why should such lopsided entries help to get a realistic model?
It turns out that these $\sigma$ entries play at least six
vital roles:

(i) In the first place, it is easy to see from the seesaw-type
formulas, that setting $\sigma = 0$ in Eq. (1) would give $m_s^0/m_b^0 
\cong (\epsilon/3)^2$ and $m_{\mu}^0/m_{\tau}^0 \cong \epsilon^2$. Thus one 
would get a ratio of ratios of 1/9 instead of the desired 
Georgi-Jarlskog value of 1/3.
If, however, $\sigma$ is present and is large compared to
$\epsilon$, then one has $m_s^0/m_b^0 \sim \sigma \epsilon/3$,
and $m_{\mu}^0/m_{\tau}^0 \sim \sigma \epsilon$, so that the
desired Georgi-Jarlskog factor emerges. 

(ii) Without the $\sigma$ entries, the
(23) blocks of $D$ and $U$ would be proportional, and consequently
there would be no KM mixing for those families, that is
$V_{cb}$ would vanish. 

(iii) Moreover, the lopsided $\sigma$ entries explain
a peculiar fact about the magnitude of $V_{cb}$. If $D_{23}$ and
$D_{32}$ were of equal magnitude, as is
the case with the purely flavor-symmetric or antisymmetric forms
usually assumed, then $V_{cb}$ would get a contribution from $D$ that
is $\sqrt{m_s/m_b}$, which is much too big (by about a factor of 4).
By having the large lopsided contribution with $\sigma > \epsilon$, one 
finds $D_{32} \ll D_{23}$, which leads to 
$V_{cb} \ll \sqrt{m_s/m_b}$, as observed. This cures one of
the main problems with textures of the Fritzsch type \cite{fritzsch} as 
noted previously by Branco and Silva-Marcos \cite{branco} in a phenomenological
study of the quark sector.  Other authors \cite{asym} have also recently 
appreciated this fact in the context of $SU(5)$ models. 

(iv) Because $\sigma$ naturally only appears in $D$ and $L$
but not in $U$, one explains why $m_c/m_t \ll m_s/m_b$ and
$m_{\mu}/m_{\tau}$, as actually observed. In the absence of the
$\sigma$ entries, one would obtain the minimal $SO(10)$
prediction $m_c^0/m_t^0 = m_s^0/m_b^0$, which is wrong by about a factor
of 20.

(v) The fact that the large $\sigma$ entry appears transposed
in $L$ relative to $D$ means that it makes the (32) mixing
of the left-handed leptons {\it large} compared to the
corresponding mixing of the left-handed quarks. One thus
explains why $\nu_{\mu}-\nu_{\tau}$ mixing is much larger than
$V_{cb}$, which is one of the great puzzles in understanding
the atmospheric neutrino oscillations.

(vi) Finally, we shall see later that the $\sigma$ entries explain why the 
$\nu_e - \nu_{\mu}$ mixing angle is significantly
smaller than $\sqrt{m_e/m_{\mu}}$, assuming the small-angle MSW
solution of the solar neutrino problem. 

What remains to be explained about the mass matrices in Eq. (1)
are the entries $\delta$ and $\delta'$. These arise very simply
from the diagrams in Fig. 3. Note that because the effective
operators are again of the form ${\bf 16}_i {\bf 16}_j 
{\bf 16}'_H {\bf 16}_H$, as for the $\sigma$ entries, these
diagrams only contribute to $D$ and $L$, and not to $U$ and $N$.
Note also that because they come from integrating out a 
${\bf 10}$ of Higgs, they are {\it flavor-symmetric}. These contributions
are assumed to be much smaller than the others.

The fact that the above terms appear and no others presupposes that
some Abelian symmetries exist that distinguish among families. Such 
symmetries have been used by many authors.  For an interesting 
recent publication with a number of references cited, see \cite{ramond}.
We note that realistic models involving
such symmetries can be constructed giving rise to the matrices in Eq. (2).
Much of the structure of such a model has already been described 
in \cite{abb2} and \cite{ab1}. In addition to what was presented there, 
there have to be $SO(10)$ ${\bf 10}$'s of Higgs fields to be integrated 
out to give the $\delta$ and $\delta'$ entries, as in Fig. 3. Such 
fields and the necessary couplings can be introduced, with appropriate 
Abelian symmetry, without destabilizing the gauge hierarchy or creating
other difficulties \cite{ab5}.

\vspace{0.5cm}

\noindent
{\bf The predictions of the matrices.}

\vspace{0.2cm}

In expressing the predictions of the model, it proves 
convenient to use instead of the parameters $\delta$ and
$\delta'$, the linear combinations $t_R \equiv \delta \sqrt{\sigma^2 + 1}/
(\sigma \epsilon/3)$ and $t_L \equiv (\delta - \sigma
\delta')/(\sigma \epsilon/3)$. Essentially, $t_L$ is the Cabibbo
angle, and $t_R$ is the corresponding mixing angle of the
{\it right-handed} $d$ and $s$ quarks. That is to say, if one diagonalizes
the (23) block of $D^0$ in Eq. (1), and calls the resulting matrix
$D^{0'}$, then $t_L \cong D^{0'}_{21}/D^{0'}_{22}$
and $t_R \cong D^{0'}_{12}/D^{0'}_{22}$. In the phase convention being
adopted, $t_R$ is real and $t_L$ has a phase that shall be called $\theta$.

The parameters, then, are $M_U/M_D$, $\sigma$, $\epsilon$, $t_R$, $\left| 
t_L \right|$, and $\theta$. In terms of these one has the following
relations, which are given only to leading order in the small
parameters $\epsilon$, $t_R$, and $t_L$:

$$\begin{array}{rclr}
m_t^0/m_b^0 & \cong & M_U/M_D, &\hfill(2a)\\
& & & \\
m_{\tau}^0/m_b^0 & \cong & 1, &\hfill(2b)\\
& & & \\
m_c^0/m_t^0 & \cong & \frac{1}{9} \epsilon^2 &\hfill(2c)\\
& & & \\
m_s^0/m_b^0 & \cong & \frac{1}{3} \epsilon \frac{\sigma}{\sigma^2 
+ 1} &\hfill(2d)\\
& & & \\
m_{\mu}^0/m_{\tau}^0 & \cong & \epsilon \frac{\sigma}{\sigma^2 
+ 1} &\hfill(2e)\\
& & & \\
m_u^0/m_c^0 & \cong & 0, &\hfill(2f)\\
& & & \\
m_d^0/m_s^0 & \cong & \left| t_L \right| t_R &\hfill(2g)\\
& & & \\
m_e^0/m_{\mu}^0 & \cong & \frac{1}{9} \left| t_L \right| t_R &\hfill(2h)\\
& & & \\
V_{cb}^0 & \cong & \frac{1}{3} \epsilon \frac{\sigma^2}{\sigma^2 + 1} 
&\hfill(2i)\\
& & & \\
V_{us}^0 & \cong & \left| t_L \right| &\hfill(2j)\\
& & & \\
V_{ub}^0 & \cong & \frac{1}{3} \epsilon \frac{1}{\sigma^2 + 1}
\left[ \sqrt{\sigma^2 + 1} t_R e^{-i \theta} - \left| t_L \right|
\right]  &\hfill(2k)\\
& & &\\
U_{\mu 3} & \equiv & \sin \theta_{\mu \tau} \cong 
- \frac{\sigma}{\sqrt{\sigma^2 + 1}} + O(\epsilon) &\hfill(2l)\\
& & &\\
U_{e2} & \cong & - \frac{1}{3} t_R \cos \theta_{\mu \tau} 
&\hfill(2m)\\
& & &\\
U_{e3} & \cong &  \frac{1}{3} t_R \sin \theta_{\mu \tau} 
&\hfill(2n)\\
\end{array}$$
\setcounter{equation}{2}

\noindent
Predictions involving only the second and third families
were given already in \cite{abb2}, where a fit was done using more exact
expressions, and taking into account renormalization-group running
from the unification scale down to low scales as well as radiative
corrections to the heavy quark masses coming from supersymmetry.
{}From $m_{\mu}/m_{\tau}$ and $V_{cb}$ which have small experimental
uncertainties, $\epsilon$ and $\sigma$, which was previously called $\rho$
in \cite{abb2}, were extracted; cf.,
Eqs. (2e) and (2i).  It was found that the best fit was given by

\begin{equation}
\begin{array}{l}
\epsilon \cong 0.139, \\
\\
\sigma \cong 1.81.
\end{array}
\end{equation}

\noindent
{}From Eq. (2c) one can then obtain a prediction for the charm quark mass.
In \cite{abb2} it was found that $m_c(m_c) = 1.10 \pm 0.1$ GeV, in satisfying
agreement with the experimental value $1.27 \pm 0.1$ GeV.

{}From Eq. (2l) one sees that the model predicts a large, and in fact a nearly
maximal, mixing of $\nu_{\mu}$ and $\nu_{\tau}$. The element $\left|
U_{\mu 3} \right|$, which
describes this mixing, receives a large contribution from diagonalizing
the mass matrix of the charged leptons, $L$. This large contribution
is well determined in the model, since $L$ is known, and comes out to
be approximately $\sigma/\sqrt{\sigma^2 + 1} \cong 0.87$. However,
$\left| U_{\mu 3} \right|$ also receives a small contribution from
diagonalizing the neutrino mass matrix, which has the usual seesaw
form $M_{\nu} = - N^T M_R^{-1} N$ \cite{gmrs}. While $N$ is known in the model, 
$M_R$ is not --- at least not without some further assumptions.
Nevertheless, one can say that the contribution to
$\left| U_{\mu 3} \right|$ from diagonalizing $M_{\nu}$ is of order
$\epsilon$, following from the form of $N$. Though relatively
small, this contribution makes impossible a precise prediction of the
$\nu_{\mu}-\nu_{\tau}$ mixing angle, though it is possible
to say that it is nearly maximal. This source
of uncertainty does not affect the predictions for the other
two lepton mixing angles, however, because of the fact that $N$,
and therefore $M_{\nu}$, have zeros in the first row and column.
These zeros are doubtless lifted by the higher order effects that
generate $m_u$, but it is easy to see that this should have
a negligible effect on $U_{e2}$ and $U_{e3}$.

The other predictions for the heavy families are that $m_b^0 \cong
m_{\tau}^0$ from Eq. (2b) and $m_s^0/m_b^0 \cong \frac{1}{3} 
m_{\mu}^0/m_{\tau}^0$ from Eqs. (2d) and (2e).
Both of these are well-known relations, and work well in the context of
supersymmetric grand unification. The foregoing four predictions were 
thoroughly discussed in \cite{abb2}. 

Let us turn now to the eight quantities that involve the first
family: $m_e$, $m_d$, $m_u$, $U_{e2}$, $U_{e3}$,
$V_{us}$, Re$(V_{ub})$, and Im$(V_{ub})$. These depend on the
three additional parameters $t_R$, $\left| t_L \right|$, and $\theta$.
The approximate magnitudes of $t_R$ and $\left| t_L \right|$ can
be estimated in a simple way, as follows. Comparing the well-known
empirical relation $\sin \theta_c \cong \sqrt{m_d/m_s}$ with
Eqs. (2j) and (2g), one sees immediately that $\left| t_L \right|
\simeq t_R \simeq \sin \theta_c$. This enables one to read off
an estimate of the Wolfenstein parameters \cite{wolf} $\rho$ and $\eta$ from
Eq. 2. To leading order in the Wolfenstein parameter $\lambda$,
one has that $\rho + i \eta = (V_{ub} V_{cs}/V_{cb} V_{us})^*$. But
the value of that invariant is given in this model (to leading
order in $\epsilon$, $t_R$, and $t_L$) by

\begin{equation}
\left( \frac{V_{ub} V_{cs}}{V_{cb} V_{us}} \right)^*
\cong \frac{1}{\sigma^2}
\left[ \sqrt{\sigma^2 + 1} \frac{t_R}{\left| t_L \right|} e^{i \theta}
-1 \right].
\end{equation}

\noindent
See Eqs. (2i), (2j), and (2k). Since $\left| t_L \right| \simeq t_R$
and $\sigma \simeq \sqrt{3}$ one has, roughly, that
$\rho + i \eta \simeq \frac{2}{3} e^{i \theta} - \frac{1}{3}$ corresponding
to a circle in the $\rho-\eta$ plane.

To get the predictions more exactly, it is best to use 
$V_{us}$ and $m_e/m_{\mu}$, 
which have the smallest experimental uncertainties, to determine 
$t_R$ and $\left| t_L \right|$; cf., Eqs. (2h) and (2j). Employing
more exact expressions than given in Eqs. (2), one finds that

\begin{equation}
\begin{array}{lcl}
\left| t_L \right| & \cong & 0.239 \\
& & \\
t_R & \cong & 0.185. \\
& & \\
\end{array}
\end{equation}

\noindent
The phase of $t_L$, denoted $\theta$, is so far undetermined.
Substituting these values into Eq. (4), and taking into account
the corrections to that equation that are higher order in
$\epsilon$, $t_R$, and $t_L$, one finds 

\begin{equation}
\rho + i \eta \cong 0.47 e^{i \theta} - 0.29. 
\end{equation}

\noindent
This circle, plotted
as the bold dashed arc in Fig. 4, slices nicely through the presently 
allowed region \cite{pdg}. 

The prediction for the $\nu_e-\nu_{\mu}$ mixing, cf. Eq. 2(m), is
that

\begin{equation}
\sin^2 2 \theta_{e \mu} \equiv 4 \left| U_{e2} \right|^2
(1 - \left| U_{e2} \right|^2) \cong 6.6 \times 10^{-3}.
\end{equation}

\noindent
This is well within the present limits $3 \times 
10^{-3} < \sin^2 2 \theta_{e \mu}
< 8 \times 10^{-3}$ 
for the small angle MSW solution to solar neutrino problem.
The crucial role of the parameter $\sigma$ should be noted. In its
absence one would find the familiar type of prediction, 
$U_{e2} \simeq \sqrt{m_e/m_{\mu}}$, which is much too large. However, 
since there is a large intermingling of the second and third
families caused by the parameter $\sigma$ --- the intermingling
that causes the large $\nu_{\mu}-\nu_{\tau}$ mixing --- the 
angles $U_{e2}$ and $U_{e3}$ get intermingled as well.
This leads to the suppression of $U_{e2}$ by a factor of
$\cos \theta_{\mu \tau} \simeq 0.7$, which brings it into the
experimentally allowed range \cite{pdg}. 

This same intermingling gives rise to a distinctive prediction of
this model, namely that

\begin{equation}
\left| U_{e3} \right| \cong \tan \theta_{\mu \tau} \left| U_{e2} \right|.
\end{equation}

\noindent
Using more exact expressions, one finds that

\begin{equation}
\sin^2 2 \theta_{e \tau} \cong 1.1 \ \tan^2 \theta_{\mu \tau} 
\sin^2 2 \theta_{e \mu}.
\end{equation}

The remaining two predictions of the model are that $m_u \approx 0$,
Eq. (2f), which has been discussed already, and that $m_d^0/m_s^0
\cong 9 m_e^0/m_{\mu^0}$, Eqs. (2g) and (2h), the other 
Georgi-Jarlskog relation, which gives $m_d/m_s \cong 1/23$ 
in excellent agreement with the current-algebra result.

A comment is in order about the robustness of these predictions.
In obtaining the value of $t_R$, we used $m_e/m_{\mu}$ on the
grounds that the experimental uncertainty in this quantity is
negligible. However, one might question whether Eq. (2h) is trustworthy. 
The point is that since some higher order effects are presumed to generate
a mass for the $u$ quark, the same effects are likely to make a small
contribution to the masses of $d$ and $e$ as well, though that is not
necessarily the case. For example, $m_u$ could arise from 
$U^0_{11} \sim 10^{-5} M_U$, and a 
comparable entry in $L$ of order $L^0_{11} \sim 10^{-5} M_D$
would affect $m_e$ at the 3\% level. This would change
$t_R$, and thus $U_{e2}$, by the same amount. The effect of such
uncertainties on the prediction for the Wolfenstein parameters 
$(\rho, \eta)$ is relatively small. The prediction in Eq. (9) is 
even more solid, since $t_R$ cancels out in
the ratio of $U_{e3}$ to $U_{e2}$. 

\vspace{0.5cm}

\noindent{\bf Conclusions}

\vspace{0.2cm}

We have presented a model of quark and lepton masses and mixings that
is based on supersymmetric $SO(10)$. The mass matrices arise
from five Yukawa terms, each of which is of a type that arises
in a straightforward way from simple tree-level diagrams.
The model contains very few parameters, so few that there are
nine predictions.

Four of the predictions follow from the basic structure of the model
and simple group-theoretic considerations. These are
$m_u \approx 0$ (which corresponds to the observed fact that
$m_u/m_t \sim 10^{-5}$, much less than $m_d/m_b \cong 10^{-3}$
and $m_e/m_{\tau} \cong 0.3 \times 10^{-3}$); $m_b^0 \cong
m_{\tau}^0$ (the well-known and successful $SU(5)$ relation);
$m_s^0/m_b^0 \cong \frac{1}{3} m_{\mu}^0/m_{\tau}^0$ (one
of the Georgi-Jarlskog relations); and $m_d^0/m_s^0
\cong 9 m_e^0/m_{\mu}^0$ (the other Georgi-Jarlskog relation).

Four other predictions are non-trivial quantitative successes of
the model. They are a prediction of the charm quark mass
that is correct within the present uncertainties; the
prediction of very large mixing between $\nu_{\mu}$ and $\nu_{\tau}$,
consistent with the recent discovery of atmospheric neutrino
oscillations; a prediction for $V_{ub}$ that slices neatly
through the middle of the allowed region for the Wolfenstein
parameters $\rho$ and $\eta$; and a prediction for 
$\sin^2 2 \theta_{e \mu}$ that is well inside the present allowed range
for the small-angle MSW solution of the solar neutrino problem.

The last prediction is of a presently unmeasured quantity: 
$\sin^2 2 \theta_{e \tau} \cong 1.1 \sin^2 2 \theta_{e \mu}$.

The model is eminently testable. The most dramatic test would
be a measurement of $\theta_{e \tau}$, which could come from improved 
proton accelerator bounds or 
perhaps from a muon collider. The next most important and fairly 
clean tests would come from a resolution of the solar neutrino
problem and the consequent precise determination of $\theta_{e \mu}$;
and from a measurement of the Wolfenstein parameters $\rho$ and $\eta$.
Much more difficult, but hopefully achievable someday, would be
precision tests of the predictions for $m_b$ and $m_c$. This would
require considerable knowledge of the spectrum of the sparticles,
knowledge of $\tan \beta$ and the Higgs masses, and better
determinations of $\alpha_s$, $m_t$, and $V_{cb}$ \cite{abb2}.\\[0.3in]

One of us (CHA) thanks the Fermilab Theoretical Physics Department for its
kind hospitality where much of his work was carried out.  The research of
SMB was supported in part by the Department of Energy under contract 
No. DE-FG02-91ER-40626.  Fermilab is operated by Universities Research
Association Inc. under contract No. DE-AC02-76CH03000 with the Department
of Energy.

\newpage
\section*{Figure Captions}

\noindent
{\bf Fig. 1:} The diagram that generates the ``$\epsilon$"
entries of the quark and lepton mass matrices. 

\vspace{0.1cm}

\noindent
{\bf Fig. 2:} The diagram that generates the ``$\sigma$"
entries of the mass matrices $L$ and $D$. 

\vspace{0.1cm}

\noindent {\bf Fig. 3:} The diagrams that produce small masses for the
first-family quarks and leptons. The family index $j$ on the spinor
field takes the values 2 or 3, giving, respectively, the $\delta$ and $\delta'$
terms of the mass matrices. Different $SO(10)$ vector Higgs, denoted with
superscript $j$, are exchanged in the two cases.

\vspace{0.1cm}

\noindent
{\bf Fig. 4} A comparison of the model with experiment for the
Wolfenstein parameters ($\rho$, $\eta$). The axes are $\rho$
and $\eta$ multiplied by the central value of 
$\left| s_{12} V_{cb} \right|$. The central
values allowed by the model lie on the bold dashed circular arc,
cf. Eq. (6). The constraints following from $|V_{ub}|,\ B$-mixing and 
$\epsilon$ extractions from experimental data are shown in the 
lightly shaded regions. The experimentally allowed region is 
indicated by the heavily shaded central region. A typical unitarity
triangle allowed by both data and the model is shown.

\newpage
\vspace*{-0.2in}
\begin{picture}(360,216)
\thicklines
\put(60,144){\vector(1,0){30}}
\put(90,144){\line(1,0){30}}
\put(120,144){\line(1,0){30}}
\put(180,144){\vector(-1,0){30}}
\put(180,144){\vector(1,0){30}}
\put(210,144){\line(1,0){30}}
\put(240,144){\line(1,0){30}}
\put(300,144){\vector(-1,0){30}}
\put(120,84){\line(0,1){12}}
\put(120,108){\vector(0,1){12}}
\put(120,132){\line(0,1){12}}
\put(180,132){\line(0,1){12}}
\put(180,120){\vector(0,-1){12}}
\put(180,84){\line(0,1){12}}
\put(240,84){\line(0,1){12}}
\put(240,108){\vector(0,1){12}}
\put(240,132){\line(0,1){12}}
\put(82,158){${\bf 16_2}$}
\put(142,158){${\bf 16}$}
\put(202,158){${\bf \overline{16}}$}
\put(262,158){${\bf 16_3}$}
\put(128,100){${\bf 10_H}$}
\put(188,100){${\bf 1_H}$}
\put(248,100){${\bf 45_H \propto B-L}$}
\put(162,45){{\bf Fig. 1}}
\end{picture}

\vspace*{-0.4in}

\begin{picture}(360,216)
\thicklines
\put(60,144){\vector(1,0){30}}
\put(90,144){\line(1,0){30}}
\put(120,144){\line(1,0){30}}
\put(180,144){\vector(-1,0){30}}
\put(180,144){\vector(1,0){30}}
\put(210,144){\line(1,0){30}}
\put(240,144){\line(1,0){30}}
\put(300,144){\vector(-1,0){30}}
\put(120,84){\line(0,1){12}}
\put(120,108){\vector(0,1){12}}
\put(120,132){\line(0,1){12}}
\put(180,132){\line(0,1){12}}
\put(180,120){\vector(0,-1){12}}
\put(180,84){\line(0,1){12}}
\put(240,84){\line(0,1){12}}
\put(240,108){\vector(0,1){12}}
\put(240,132){\line(0,1){12}}
\put(78,158){${\bf \overline{5}(16_2)}$}
\put(142,158){${\bf 5(10)}$}   
\put(202,158){${\bf \overline{5}(10)}$}
\put(262,158){${\bf 10(16_3)}$}  
\put(125,100){${\bf 1(16_H)}$}
\put(188,100){${\bf 1_H}$}
\put(248,100){${\bf \overline{5}(16'_H)}$}
\put(162,45){{\bf Fig. 2}}
\end{picture}

\vspace*{-0.2in}

\begin{picture}(360,216)
\thicklines
\put(90,180){\vector(1,0){45}}
\put(135,180){\line(1,0){45}}
\put(270,180){\vector(-1,0){45}}
\put(180,180){\line(1,0){45}}
\put(180,120){\line(0,1){12}}
\put(180,144){\line(0,1){12}}
\put(180,168){\line(0,1){12}}
\put(127,194){${\bf 16_1}$}
\put(217,194){${\bf 16_j}$}
\put(190,148){${\bf 10^j_H}$}
\put(96,120){\line(1,0){12}}
\put(120,120){\vector(1,0){12}}
\put(144,120){\line(1,0){12}}
\put(168,120){\line(1,0){12}}
\put(180,120){\line(1,0){12}}
\put(204,120){\line(1,0){12}}
\put(240,120){\vector(-1,0){12}}
\put(252,120){\line(1,0){12}}
\put(124,98){${\bf 16_H}$}
\put(208,98){${\bf 16'_H}$}
\put(162,60){{\bf Fig. 3}}
\end{picture}
\newpage
\begin{figure}
\epsfxsize=1.0\hsize
\epsfxsize=1.0\hsize
\epsfbox{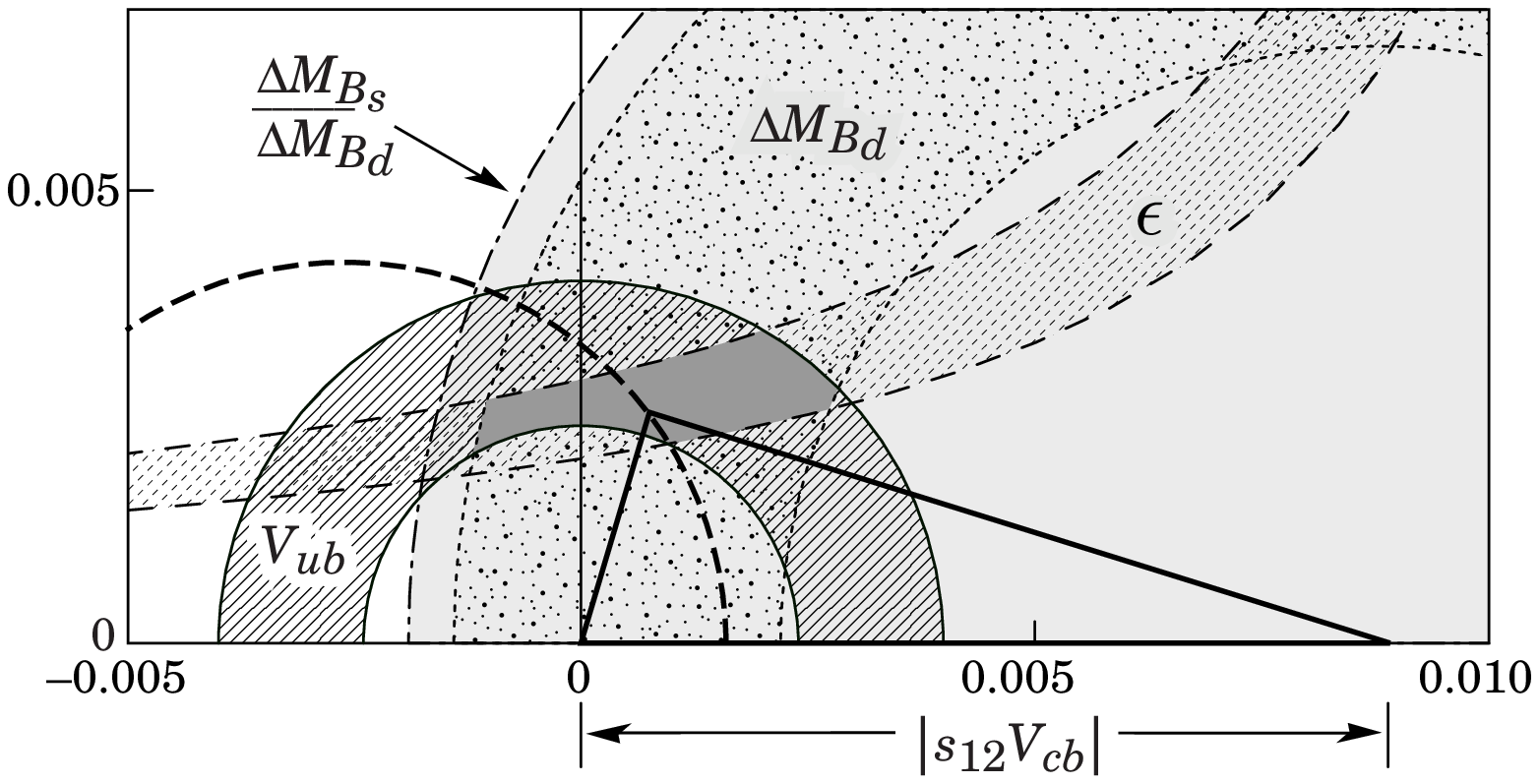}
\end{figure}
\nopagebreak
\begin{center}
        Fig. 4
\end{center}
\end{document}